\documentclass[a4paper,12pt,fleqn]{article}

\usepackage[tbtags]{amsmath}

\newcommand{\ud}{\mathrm{d}}
\renewcommand{\vec}{\mathbf}

\title{\bf New approach to Hamiltonian formulation of Hall magnetohydrodynamics}

\author{{\normalsize \textbf{Kuldeep Kumar}}\\
{\small \textit{Department of Physics, Panjab University, Chandigarh 160014, India}}\\[-0.5ex]
{\small \texttt{kuldeepk@pu.ac.in}}
}

\date{}

\begin{document}

\maketitle

\begin{abstract}
We present a new Hamiltonian formulation of barotropic Hall magnetohydrodynamics in two complementary approaches based on Dirac's constraint analysis. In one case the Hamiltonian is canonical involving physical variables only but the brackets are noncanonical, while in the other case the phase space is enlarged to retain the canonical structure of brackets.
\end{abstract}


\section{Introduction}

There are two descriptions to study fluid theories. One is the Lagrange description which focuses on the coordinates of individual fluid particles. This description has the advantage that one can write down the Hamiltonian in terms of coordinates of fluid particles and their momenta and the equations of motion follow in the usual way by bracketing the coordinates and momenta with the Hamiltonian and using canonical (Poisson) brackets between coordinates and momenta. The other description is the Euler description which is akin to classical field theory in physical spacetime. In this description, which is more elegant, one observes field quantities at a fixed point rather than following a given fluid element. However the Euler variables are noncanonical and hence the brackets which correctly reproduce the equations of motion in Hamiltonian formulation are also noncanonical.

Magnetohydrodynamics (MHD) is concerned with electrically conducting fluids and with the effects arising through the interaction of fluid motion and any ambient magnetic field that may be present. An action principle for ideal MHD was originally given by Newcomb \cite{Newcomb1962} and later followed by others \cite{Lundgren1963, Calkin1963, PH1966, Merches1969, BO2000, ZQBB2014, Y2015}. The noncanonical Hamiltonian formulation of ideal MHD was given in \cite{MG1980, Morrison1982}. To arrive at these brackets one can start with the Lagrange description first and use its mapping to the Euler version \cite{Morrison2009}. Alternatively, one can choose a suitable Clebsch form for velocity (and optionally for magnetic field) \cite{Morrison1982, Zakharov1971, Morrison1998}, identify the canonical pairs of variables and then map to the noncanonical brackets for physical variables. A new approach to obtain the noncanonical brackets starting from an action principle and following Dirac's constraint analysis \cite{Dirac1964} was presented in \cite{BK2016}.

One of the important extensions of ideal MHD is the Hall MHD where one accounts for the difference between the motion of the two species in a typical plasma. An action principle and a Hamiltonian formulation for Hall MHD in terms of Clebsch potentials were given in \cite{YH2013} where noncanonical brackets necessary to reproduce the equations of motion were posited. A similar Hamiltonian formalism for extended MHD was carried out in \cite{AKY2015}. These noncanonical brackets were derived through Euler-Lagrange map in \cite{DML2016}. However, a systematic derivation of these noncanonical brackets solely within the Euler framework is lacking. To fill this gap, here we provide a new Hamiltonian formulation of Hall MHD based on Dirac's constraint analysis \cite{Dirac1964}, which is an extension of our earlier approach \cite{BK2016} in the context of ideal MHD.

We present a brief review of Hall MHD in Sec.~\ref{hmhdrev} which also helps to set up notations. In Sec.~\ref{hmhdham} we give Hamiltonian formulation of Hall MHD in two complimentary approaches. In the first part we obtain the noncanonical brackets as Dirac brackets. In the second part we enlarge the phase space by changing the Hamiltonian to total Hamiltonian while the brackets retain their canonical nature. We summarise our results in Sec.~\ref{conclu}.


\section{\label{hmhdrev}Brief review of Hall MHD}

In the Euler description, MHD is described in terms of the fluid density
$\rho(\vec{x},t)$, the fluid velocity $\vec{v}(\vec{x},t)$, entropy per unit mass $s(\vec{x},t)$ and the magnetic field $\vec{B}(\vec{x},t)$. Basic Hall MHD equations for barotropic plasma are\footnote{Notation: $\partial_0=\partial/\partial t$, $\partial_i=\partial/\partial x_i\; i=1,2,3$, summation over repeated index implied, SI units for electrodynamics. For the general (non-barotropic) case, equation \eqref{eqB} will have an additional term on the right-hand side proportional to $\vec{\nabla}\times (\frac{1}{\rho}\vec{\nabla}p_\mathrm{e})$,  $p_\mathrm{e}$ being the electron pressure. Such a term vanishes for the barotropic case in which $p_\mathrm{e}$ is a function of $\rho$ alone (and not of $s$).
}
\begin{gather}
\label{eqrho} \partial_0 \rho + \vec{\nabla} \cdot (\rho \vec{v}) = 0, \\
\label{eqs} \partial_0 s + \vec{v} \cdot \vec{\nabla} s = 0, \\
\label{eqB} \partial_0 \vec{B} = \vec{\nabla} \times (\vec{v} \times \vec{B}) - \frac{m}{\mu e} \vec{\nabla} \times \left[ \frac{(\vec{\nabla}\times \vec{B}) \times \vec{B}}{\rho} \right], \\
\label{eqv} \partial_0 \vec{v} + (\vec{v}\cdot \vec{\nabla}) \vec{v} = -\frac{1}{\rho} \vec{\nabla} p + \frac{1}{\mu \rho} (\vec{\nabla}\times \vec{B}) \times \vec{B},
\end{gather}
where $m$ is the ion mass, $e$ the electron charge and $p(\vec{x},t)= \rho^2 {\partial \epsilon}/{\partial \rho}$ the fluid pressure, $\epsilon(\rho,s)$ being the thermodynamic internal energy per unit mass.
Magnetic field also satisfies the Gauss's law
\begin{equation}
\label{gaussB} \vec{\nabla} \cdot \vec{B} = 0.
\end{equation}
For the ideal MHD case the second term on the right-hand side of \eqref{eqB} will be absent.

Equations \eqref{eqrho}--\eqref{eqv} follow from the Hamiltonian
\begin{equation}\label{hammhd3}
H = \int\!\ud^3 x \Big( \frac{1}{2} \rho v^2 + \rho \epsilon(\rho,s) + \frac{B^2}{2\mu} \Big)
\end{equation}
and the following nonvanishing brackets \cite{DML2016}
\begin{gather}
\label{bravrho} \{v_i(\vec{x}), \rho(\vec{x}')\} = -\partial_i \delta(\vec{x}-\vec{x}'), \\
\label{bravs} \{v_i(\vec{x}), s(\vec{x}')\} = \frac{\partial_i s}{\rho} \delta(\vec{x}-\vec{x}'), \\
\label{braBv} \{B_i(\vec{x}), v_j(\vec{x}')\} = \delta_{ij} \left( \frac{B_k}{\rho} \right)_{x'} \partial_k \delta(\vec{x}-\vec{x}') - \left( \frac{B_i}{\rho} \right)_{x'} \partial_j \delta(\vec{x}-\vec{x}'),\\
\label{bravv} \{v_i(\vec{x}), v_j(\vec{x}')\} = \frac{\omega_{ij}}{\rho} \delta(\vec{x}-\vec{x}'),\\
\label{braBB} \{ B_i (\vec{x}), B_j (\vec{x}')\} = \frac{m}{e} \varepsilon_{kln} \varepsilon_{lrj}\varepsilon_{nqi} \partial_q \partial'_r \left[\frac{B_k}{\rho}\delta(\vec{x}-\vec{x}')\right],
\end{gather}
where $\omega_{ij}=\partial_i v_j - \partial_j v_i$ is the vorticity. The only difference from the ideal MHD case is that here we have a nonzero $B_i$--$B_j$ bracket, which is necessary to reproduce \eqref{eqB}.

Action for barotropic Hall MHD can be written as \cite{YH2013}
\begin{gather}
\label{acthmhd}
S = \int\!\ud t\, \ud^3 x \Big( -\theta \partial_0 \rho - \lambda\rho \partial_0 s - \alpha\rho \partial_0 \beta + \frac{e}{m}\rho \psi \partial_0 \phi - \frac{1}{2} \rho v^2 - \rho \epsilon(\rho,s) - \frac{B^2}{2\mu} \Big),\\
\label{eqvcd} v_i = -\partial_i \theta + \lambda\partial_i s + \alpha\partial_i \beta -\frac{e}{m} \psi\partial_i \phi,\\
\label{eqBcd} \vec{B} = \vec{\nabla}\psi \times \vec{\nabla}\phi,
\end{gather}
where we have chosen a Clebsch-type decomposition for both the velocity and the magnetic field.\footnote{The formulation in \cite{YH2013} does not include the additional field, subjected to Lin's constraint \cite{Lin1963}, necessary to include vortical flows, which we have included here. Thus we have 2 additional variables, $\alpha$ and $\beta$, in our action as compared to that of  \cite{YH2013}.}  The basic fields of the action are $\rho$, $\theta$, $s$, $\lambda$, $\alpha$, $\beta$, $\psi$ and $\phi$ only. First we briefly demonstrate how Eqs.~\eqref{eqrho}--\eqref{eqv} follow from the action \eqref{acthmhd} as the Euler-Lagrange equations.

Variations of the action \eqref{acthmhd} with respect to $\theta$ and $\lambda$ give Eqs.~\eqref{eqrho} and \eqref{eqs} respectively. Similarly, varying the action with respect to $\alpha$, $\beta$, $s$, $\psi$ and $\phi$ and using Eq.~\eqref{eqrho} give the following equations, respectively:
\begin{gather}
\label{eqbeta} \partial_0 \beta + v_i \partial_i \beta = 0,\\
\label{eqalpha} \partial_0 \alpha + v_i \partial_i \alpha = 0,\\
\label{eqlambda} \partial_0 \lambda + v_i \partial_i \lambda - \frac{\partial \epsilon}{\partial s} = 0,\\
\label{eqphi} \partial_0 \phi + v_i \partial_i\phi -\frac{m}{\mu e \rho} (\vec{\nabla}\times \vec{B})\cdot\vec{\nabla}\phi = 0,\\
\label{eqpsi} \partial_0 \psi + v_i \partial_i\psi -\frac{m}{\mu e \rho} (\vec{\nabla}\times \vec{B})\cdot\vec{\nabla}\psi = 0.
\end{gather}
Finally, variation with respect to $\rho$ yields
\begin{equation*}
\label{eqtheta0} \partial_0 \theta -\lambda \partial_0 s - \alpha \partial_0 \beta + \frac{e}{m}\psi \partial_0 \phi  - \frac{v^2}{2} - \epsilon - \rho \frac{\partial \epsilon}{\partial \rho} = 0,
\end{equation*}
which using \eqref{eqs}, \eqref{eqbeta}, \eqref{eqphi} and \eqref{eqvcd} reduces to
\begin{equation}
\label{eqtheta} \partial_0 \theta + v_i \partial_i \theta + \frac{v^2}{2} - \epsilon - \rho \frac{\partial \epsilon}{\partial \rho} + \frac{1}{\mu \rho} \psi (\vec{\nabla}\times \vec{B})\cdot\vec{\nabla}\phi = 0.
\end{equation}
Now equation \eqref{eqB} is reproduced as follows. From \eqref{eqBcd}, we have
\begin{equation}
\begin{split}
\partial_0 \vec{B} = \partial_0 (\vec{\nabla}\psi \times \vec{\nabla}\phi) = \vec{\nabla}\times (\partial_0 {\psi}\, \vec{\nabla}\phi - \partial_0 {\phi}\, \vec{\nabla}\psi).
\end{split}
\end{equation}
Substituting for $\partial_0 \phi$ and $\partial_0 \psi$ from Eqs.~\eqref{eqphi} and \eqref{eqpsi} and finally again using \eqref{eqBcd} to express the result in terms of $\vec{B}$ yields Eq.~\eqref{eqB}. For reproducing Eq.~\eqref{eqv}, we act $(\partial_0 + \vec{v}\cdot\vec{\nabla})$ on Eq.~\eqref{eqvcd},
\begin{equation}
(\partial_0 + \vec{v}\cdot\vec{\nabla}) v_i = (\partial_0 + \vec{v}\cdot\vec{\nabla})(-\partial_i \theta + \lambda\partial_i s + \alpha\partial_i \beta -\frac{e}{m} \psi\partial_i \phi),
\end{equation}
and eliminate the time-derivatives on the right-hand side using Eqs.~\eqref{eqrho}, \eqref{eqs}, \eqref{eqbeta}--\eqref{eqtheta}. Then using Eq.~\eqref{eqBcd} and
\begin{equation}\label{x1}
-\partial_i \Big( \epsilon + \rho \frac{\partial \epsilon}{\partial\rho}\Big) + \frac{\partial\epsilon}{\partial s}\partial_i s = -\frac{1}{\rho} \partial_i p,
\end{equation}
which follows from the definition of pressure, $p = \rho^2 {\partial \epsilon}/{\partial \rho}$, Eq.~\eqref{eqv} is reproduced. The Gauss's law \eqref{gaussB} is inbuilt in this formalism, this is ensured by Eq.~\eqref{eqBcd}.


\section{\label{hmhdham}Hamiltonian formulation of barotropic Hall MHD}

Extending our earlier approach \cite{BK2016} in the context of ideal MHD, we present here a new Hamiltonian formulation of Hall MHD based on Dirac's constraint analysis \cite{Dirac1964}. For the action \eqref{acthmhd} we write down the momenta conjugate to $\rho$, $s$, $\beta$, $\phi$, $\theta$, $\lambda$, $\alpha$ and $\psi$ as
\begin{equation}\label{momenta}
\begin{aligned}
&\pi_\rho = -\theta, \quad \pi_s = -\lambda\rho, \quad \pi_\beta = -\alpha\rho, \quad \pi_\phi = \frac{e}{m}\rho\psi, \\
&\pi_\theta = 0, \quad \pi_\lambda = 0, \quad \pi_\alpha = 0, \quad \pi_\psi = 0, 
\end{aligned}
\end{equation}
respectively. The canonical Hamiltonian is
\begin{equation}\label{hamhmhd}
H = \int\!\ud^3 x\, \mathcal{H} = \int\!\ud^3 x \Big( \frac{1}{2} \rho v^2(\rho,s,\theta,\lambda,\alpha,\beta,\psi,\phi) + \rho \epsilon(\rho,s) + \frac{B^2 (\psi,\phi)}{2\mu} \Big).
\end{equation}
We label the primary constraints of the theory following from \eqref{momenta} as
\begin{equation}\label{cons}
\begin{aligned}
&\Omega_1 = \pi_\rho + \theta, \quad \Omega_2 = \pi_s + \lambda\rho, \quad \Omega_3 = \pi_\beta + \alpha\rho, \quad
\Omega_{4} = \pi_\phi - \frac{e}{m}\rho\psi,\\
&\Omega_5 = \pi_\theta, \quad \Omega_6 = \pi_\lambda, \quad \Omega_7 = \pi_\alpha, \quad \Omega_{8} = \pi_\psi.
\end{aligned}
\end{equation}
All these 8 constraints are second-class as seen from their Poisson brackets. We now construct the constraint matrix of the Poisson brackets,\footnote{All brackets are equal-time, so time argument is omitted for convenience.}
\begin{equation}
\Lambda_{a,b}(\vec{x},\vec{x}') = \{\Omega_a(\vec{x}), \Omega_b(\vec{x}')\},\quad a,b=1,\ldots,8,
\end{equation}
which has the following nonvanishing components:
\begin{equation}
\begin{aligned}
&\Lambda_{1,2}(\vec{x},\vec{x}') = -\lambda \delta(\vec{x}-\vec{x}'), \quad
\Lambda_{1,3}(\vec{x},\vec{x}') = -\alpha \delta(\vec{x}-\vec{x}'), \\
&\Lambda_{1,4}(\vec{x},\vec{x}') = \frac{e}{m}\psi \delta(\vec{x}-\vec{x}'),\quad
\Lambda_{1,5}(\vec{x},\vec{x}') =  \delta(\vec{x}-\vec{x}'), \\
&\Lambda_{2,6}(\vec{x},\vec{x}') = \Lambda_{3,7}(\vec{x},\vec{x}') = \rho \delta(\vec{x}-\vec{x}'),\quad
\Lambda_{4,8}(\vec{x},\vec{x}') = -\frac{e}{m}\rho \delta(\vec{x}-\vec{x}').
\end{aligned}
\end{equation}
The inverse matrix, $\Lambda^{-1}(\vec{x},\vec{x}')$, which is defined as
\begin{equation}
\int\!\ud y^3 \Lambda^{-1}_{a,b}(\vec{x},\vec{y}) \Lambda_{b,c}(\vec{y},\vec{x}') = \delta_{ac} \delta(\vec{x}-\vec{x}'),
\end{equation}
has the following nonvanishing components:
\begin{equation}
\begin{aligned}
&\Lambda^{-1}_{1,5}(\vec{x},\vec{x}') = -\delta(\vec{x}-\vec{x}'), \quad
\Lambda^{-1}_{2,6}(\vec{x},\vec{x}') = \Lambda^{-1}_{3,7}(\vec{x},\vec{x}') = -\frac{1}{\rho} \delta(\vec{x}-\vec{x}'), \\
&\Lambda^{-1}_{4,8}(\vec{x},\vec{x}') = \frac{m}{e\rho} \delta(\vec{x}-\vec{x}'), \quad
\Lambda^{-1}_{5,6}(\vec{x},\vec{x}') = -\frac{\lambda}{\rho} \delta(\vec{x}-\vec{x}'),\\
&\Lambda^{-1}_{5,7}(\vec{x},\vec{x}') = -\frac{\alpha}{\rho} \delta(\vec{x}-\vec{x}'), \quad
\Lambda^{-1}_{5,8}(\vec{x},\vec{x}') = -\frac{\psi}{\rho} \delta(\vec{x}-\vec{x}').
\end{aligned}
\end{equation}
The canonical Hamiltonian \eqref{hamhmhd} is replaced, in Dirac's procedure, by the total Hamiltonian
\begin{equation}\label{hamthmhd}
\begin{split}
H_\mathrm{T} &= \int\!\ud^3 x \Big( \mathcal{H} + C_a \Omega_a\Big)\\
&= \int\!\ud^3 x \Big( \frac{1}{2} \rho v^2(\rho,s,\theta,\lambda,\alpha,\beta,\psi,\phi) + \rho \epsilon(\rho,s) + \frac{B^2 (\psi,\phi)}{2\mu} + C_a \Omega_a\Big),
\end{split}
\end{equation}
where $C_a$, $a=1,\ldots, 8$, are the Lagrange multiplier fields implementing the constraints \eqref{cons}. Now we have two options. We may eliminate the constraints by working with Dirac brackets instead of Poisson brackets. This will give a formulation where the Hamiltonian retains its canonical structure \eqref{hamhmhd} but the basic algebra is given by the Dirac brackets. In the other option we may fix the multipliers in \eqref{hamthmhd} by requiring time-conservation of the constraints. Then we have a formulation involving the total (noncanonical) Hamiltonian \eqref{hamthmhd} but all brackets are canonical.

\subsection{Hamiltonian formulation in terms of noncanonical brackets}

The second-class constraints (\ref{cons}) can be eliminated by computing the Dirac brackets, denoted by a star, which are defined in terms of the usual canonical (Poisson) brackets as
\begin{equation}\label{new}
\begin{split}
\{F(\vec{x}), G(\vec{x}')\}^* &= \{F(\vec{x}), G(\vec{x}')\}\\
&{}\quad - \int\!\ud^3 y\, \ud^3 z\, \{F(\vec{x}), \Omega_a(\vec{y})\} \Lambda^{-1}_{a,b}(\vec{y},\vec{z}) \{\Omega_b(\vec{z}), G(\vec{x}')\}.
\end{split}
\end{equation}
The nonvanishing Dirac brackets among various fields, in our case, turn out to be
\begin{equation}\label{brac}
\begin{aligned}
&\{\rho(\vec{x}), \theta(\vec{x}')\}^* = - \delta(\vec{x}-\vec{x}'), \quad
\{\lambda(\vec{x}), \theta(\vec{x}')\}^* = \frac{\lambda}{\rho} \delta(\vec{x}-\vec{x}'), \\
&\{\alpha(\vec{x}), \theta(\vec{x}')\}^* = \frac{\alpha}{\rho} \delta(\vec{x}-\vec{x}'), \quad
\{\psi(\vec{x}), \theta(\vec{x}')\}^* = \frac{\psi}{\rho} \delta(\vec{x}-\vec{x}'), \\
&\{\lambda(\vec{x}), s(\vec{x}')\}^* = \{\alpha(\vec{x}), \beta(\vec{x}')\}^* = \frac{1}{\rho} \delta(\vec{x}-\vec{x}'), \\
&\{\psi(\vec{x}), \phi(\vec{x}')\}^* = -\frac{m}{e\rho} \delta(\vec{x}-\vec{x}').
\end{aligned}
\end{equation}
The physical fields $\rho$ and $s$ have vanishing brackets among themselves.

Let us make a consistency check at this stage. From the brackets \eqref{brac} one can easily identify the canonical pairs, which are $(\theta,\rho)$, $(\rho\lambda,s)$, $(\rho\alpha, \beta)$ and $(\phi,e\rho\psi/m)$.  The same set of pairs can also be identified from the action \eqref{acthmhd} itself.

The constraints \eqref{cons} can be now eliminated by working with the Dirac brackets. The total Hamiltonian then reduces to the canonical form \eqref{hamhmhd}. It is now advantageous to obtain the Dirac brackets among the physical fields, $\rho$, $s$, $v_i$ and $B_i$, because then we can use the Hamiltonian \eqref{hamhmhd} explicitly expressed in terms of these physical fields,
\begin{equation}\label{hamhmhd2}
H = \int\!\ud^3 x \Big( \frac{1}{2} \rho v^2 + \rho \epsilon(\rho,s) + \frac{B^2}{2\mu} \Big),
\end{equation}
to obtain the equations for barotropic Hall MHD. As mentioned earlier, $\rho$ and $s$ have vanishing brackets among themselves. Using brackets \eqref{brac} and Eq.~\eqref{eqvcd} it is straightforward to see that
\begin{gather}
\label{dbravrho} \{v_i(\vec{x}), \rho(\vec{x}')\}^* = -\partial_i \delta(\vec{x}-\vec{x}'), \\
\label{dbravs} \{v_i(\vec{x}), s(\vec{x}')\}^* = \frac{\partial_i s}{\rho} \delta(\vec{x}-\vec{x}').
\end{gather}
Now we compute the $B_i$--$B_j$ bracket using \eqref{eqBcd} and the $\psi$--$\phi$ bracket in \eqref{brac}, which turns out to be
\begin{equation}
\{ B_i (\vec{x}), B_j (\vec{x}')\}^* = \frac{m}{e} \varepsilon_{lrj}\varepsilon_{nqi} \partial_q \partial'_r \left[\frac{1}{\rho}(\partial_l \psi \partial_n \phi - \partial_n \psi \partial_l \phi)\delta(\vec{x}-\vec{x}')\right]
\end{equation}
Once we notice that $\partial_l \psi \partial_n \phi - \partial_n \psi \partial_l \phi = \varepsilon_{lnk}B_k$, which follows from \eqref{eqBcd}, this reduces to
\begin{equation}\label{dbraBB}
\{ B_i (\vec{x}), B_j (\vec{x}')\}^* = \frac{m}{e} \varepsilon_{kln} \varepsilon_{lrj}\varepsilon_{nqi} \partial_q \partial'_r \left[\frac{B_k}{\rho}\delta(\vec{x}-\vec{x}')\right],
\end{equation}
which is precisely the bracket \eqref{braBB}. For the computation of $B_i$--$v_j$ and $v_i$--$v_j$ brackets, it is convenient to work out some intermediate steps first. Use of brackets \eqref{brac} and Eq.~\eqref{eqvcd} yields
\begin{gather}
\label{bravtheta} \{v_i(\vec{x}), \theta(\vec{x}')\}^* = \frac{1}{\rho}\left( \lambda \partial_i s + \alpha \partial_i \beta - \frac{e}{m}\psi \partial_i \phi \right) \delta(\vec{x}-\vec{x}'), \\
\label{bravlambda} \{v_i(\vec{x}), \lambda(\vec{x}')\}^* = \frac{1}{\rho} \partial_i \lambda\, \delta(\vec{x}-\vec{x}'), \\
\label{bravalpha} \{v_i(\vec{x}), \alpha(\vec{x}')\}^* = \frac{1}{\rho} \partial_i \alpha\, \delta(\vec{x}-\vec{x}'), \\
\label{bravbeta} \{v_i(\vec{x}), \beta(\vec{x}')\}^* = \frac{1}{\rho}\partial_i  \beta\, \delta(\vec{x}-\vec{x}'), \\
\label{bravpsi} \{v_i(\vec{x}), \psi(\vec{x}')\}^* = \frac{1}{\rho}\partial_i  \psi\, \delta(\vec{x}-\vec{x}'), \\
\label{bravphi} \{v_i(\vec{x}), \phi(\vec{x}')\}^* = \frac{1}{\rho}\partial_i  \phi\, \delta(\vec{x}-\vec{x}').
\end{gather}
Equation \eqref{eqvcd} also gives the following expression for vorticity in terms of basic fields:
\begin{equation}\label{vort}
\begin{split}
\omega_{ij} &= \partial_i v_j - \partial_j v_i \\
&= \partial_i \lambda \partial_j s + \partial_i \alpha \partial_j \beta - \frac{e}{m} \partial_i \psi \partial_j \phi -\langle i\leftrightarrow j\rangle,
\end{split}
\end{equation}
where $\langle i\leftrightarrow j\rangle$ stands for the previous terms with $i$ and $j$ interchanged.

Now we proceed to evaluate the $B_i$--$v_j$ and $v_i$--$v_j$ brackets:
\begin{gather}
\{B_i(\vec{x}), v_j(\vec{x}')\}^* = \left\{[\varepsilon_{ikl}\partial_k \psi \partial_l \phi](\vec{x}), v_j(\vec{x}') \right\}^*, \\
\{v_i(\vec{x}), v_j(\vec{x}')\}^* = \left\{v_i(\vec{x}), \Big[-\partial_j \theta + \lambda\partial_j s + \alpha\partial_j \beta - \frac{e}{m}\psi \partial_j \phi \Big](\vec{x}')\right\}^*.
\end{gather}
Brackets \eqref{dbravrho}, \eqref{dbravs}, \eqref{bravtheta}--\eqref{bravphi} are used to simplify the right-hand sides of the above equations and finally we use \eqref{eqBcd} and \eqref{vort} to express the result in terms of physical fields. Then we get
\begin{gather}
\label{dbraBv} \{B_i(\vec{x}), v_j(\vec{x}')\}^* = \delta_{ij} \left( \frac{B_k}{\rho} \right)_{x'} \partial_k \delta(\vec{x}-\vec{x}') - \left( \frac{B_i}{\rho} \right)_{x'} \partial_j \delta(\vec{x}-\vec{x}'),\\
\label{dbravv} \{v_i(\vec{x}), v_j(\vec{x}')\}^* = \frac{\omega_{ij}}{\rho} \delta(\vec{x}-\vec{x}'),
\end{gather}
which are precisely the brackets \eqref{braBv} and \eqref{bravv}.

From these noncanonical (Dirac) brackets and the Hamiltonian \eqref{hamhmhd2}, the barotropic Hall MHD equations \eqref{eqrho}--\eqref{eqv} follow in the usual way: $\partial_0 \rho = \{\rho, H\}^*$, $\partial_0 s = \{s, H\}^*$, $\partial_0 B_i = \{B_i, H\}^*$, $\partial_0 v_i = \{v_i, H\}^*$.

\subsection{Hamiltonian formulation in terms of canonical brackets}

We now discuss the second option. This will involve the total (noncanonical) Hamiltonian but all the brackets will be canonical. For that we fix the multipliers $C_a$ appearing in \eqref{hamthmhd}. Conserving all the primary constraints with time,
\begin{equation}
\partial_0 \Omega_a = \left\{ \Omega_a, H_\mathrm{T}\right\} = 0,
\end{equation}
gives conditions on $C_a$. We get 8 such conditions in total corresponding to the 8 constraints $\Omega_a$. These conditions uniquely fix the multipliers:
\begin{equation}\label{Cs}
\begin{aligned}
&C_1 = -\partial_i (\rho v_i), \quad
C_2 =  -v_i \partial_i s, \quad
C_3 = -v_i \partial_i \beta, \\
&C_{4} = -v_i \partial_i \phi - \frac{m}{\mu e \rho} \varepsilon_{ikl} \partial_k B_i \partial_l \phi,\\
&C_5 = -\frac{v^2}{2} - v_i \partial_i \theta + \epsilon + \rho \frac{\partial \epsilon}{\partial \rho} + \frac{\psi}{\mu \rho} \varepsilon_{ikl}  \partial_k B_i \partial_l \phi, \\
&C_6 = -v_i \partial_i \lambda + \frac{\partial \epsilon}{\partial s}, \quad
C_7 = -v_i \partial_i \alpha, \\
&C_{8} = -v_i \partial_i \psi - \frac{m}{\mu e \rho} \varepsilon_{ikl} \partial_k B_i \partial_l \psi,
\end{aligned}
\end{equation}
where $v_i$ and $B_i$ appearing on the right-hand sides of these equations are expressed in terms of basic fields as given in \eqref{eqvcd} and \eqref{eqBcd}. Equations of motion now follow using the standard Poisson brackets ($\{ \rho(\vec{x}),\pi_\rho (\vec{x}')\} = \delta(\vec{x}-\vec{x}')$, etc.)\ and the Hamiltonian $H_\mathrm{T}$ given in \eqref{hamthmhd} with the multipliers $C_a$ as given in \eqref{Cs}.  Equations \eqref{eqrho} and \eqref{eqs} are obtained in a straightforward manner. Derivation of Eqs.~\eqref{eqB} and \eqref{eqv} is a bit involved, which we now explicitly demonstrate.

Using \eqref{eqvcd} and \eqref{eqBcd} for $v_i$ and $B_i$, the Hamiltonian $H_\mathrm{T}$ given in \eqref{hamthmhd} and the standard Poisson brackets, we get
\begin{gather}
\partial_0 B_i = \left\{ B_i, H_\mathrm{T} \right\} = \varepsilon_{ikl} (\partial_k \psi \partial_l C_4 - \partial_k \phi \partial_l C_8), \\
\begin{split}
\partial_0 v_i &= \left\{ v_i, H_\mathrm{T} \right\} \\
&= -\partial_i C_5 + C_6 \partial_i s + \lambda \partial_i C_2 + C_7 \partial_i \beta + \alpha \partial_i C_3 - \frac{e}{m} C_8 \partial_i \phi - \frac{e}{m} \psi \partial_i C_4.
\end{split}
\end{gather}
Now we substitute the multipliers from \eqref{Cs}, use \eqref{x1} and finally use \eqref{eqvcd} and \eqref{eqBcd} to express the right-hand sides in terms of physical fields. Then we get
Eqs.~\eqref{eqB} and \eqref{eqv}.

Thus, the barotropic Hall MHD equations can be obtained either from a canonical Hamiltonian \eqref{hamhmhd2} using the noncanonical brackets \eqref{dbravrho}, \eqref{dbravs}, \eqref{dbraBB}, \eqref{dbraBv} and \eqref{dbravv}, or from a noncanonical Hamiltonian \eqref{hamthmhd}, with $C_a$ as given in \eqref{Cs}, using the canonical brackets.


\section{\label{conclu}Conclusions}

We have presented a new Hamiltonian formulation of barotropic Hall MHD in Euler variables based on Dirac's constraint analysis, which is an extension of our earlier approach \cite{BK2016} in the context of ideal MHD.

We started with the action considered previously in \cite{YH2013}. But we have included an additional term also to incorporate Lin's constraint, so we have 2 additional variables in the action as compared to that of \cite{YH2013}. Following Dirac's constraint method, this system turned out to second-class as all the constraints were found to be second-class. Dirac brackets were constructed to eliminate the constraints. These brackets were just the noncanonical brackets posited in \cite{YH2013}. Within this formulation we presented a complimentary viewpoint also where the constraints were not eliminated but rather implemented by Lagrange multipliers in the construction of total Hamiltonian. The Lagrange multipliers were fixed by requiring time-conservation of the constraints. Then we had a Hamiltonian formulation where the brackets were canonical in an enlarged phase space and the equations of motion followed by using these canonical (Poisson) brackets and the new (total) Hamiltonian.

It is pertinent to note that the noncanonical brackets for Hall MHD were earlier obtained in \cite{DML2016} using Lagrange to Euler map. However a systematic derivation of these brackets solely within the Euler description was lacking. We have provided just that. Moreover, we have given an alternative Hamiltonian formulation also, which is completely new, in terms of canonical brackets and a modified (total) Hamiltonian. Essentially, it is a trade-off between a canonical Hamiltonian with noncanonical brackets and a noncanonical Hamiltonian with canonical brackets.


\section*{Acknowledgement}

Author would like to thank Rabin Banerjee for helpful discussions.



\end{document}